# Possible strain-induced directional superconductivity in graphene

**Bumned Soodchomshom**

Department of Physics, Faculty of Science, Kasetsart University, Bangkok 10900, Thailand

**Abstract**
    Applying large strain in zigzag direction, gapless graphene may turns into gapped graphene at the critical strain. The energy gap between valence and conduction bands is created above the critical deformation. We theoretically predict that, using the Landauer formalism to study conductance in ballistic limit, the strain dependence of ballistic conductance G, related to tight-binding-based carriers, evolves into a tremendously large conductance $G \to \infty$ at the critical strain, found only for the conductance of current along armchair direction. This directional superconductance may lead graphene to resemble a superconductor. The strain-induced energy gap plays the role of the superconducting gap. This behavior is due to the fact that strain-induced change of electronic properties leads to highly anisotropic fermions to cause this tremendously large conductance.

**Keywords:** graphene; strain; superconductor; phase transition

## 1. Introduction

    Graphene or a monolayer of graphite, an extremely thin sheet in order of one carbon-atomic thick, has first fabricated in the recent year [1]. After its discovery, graphene has become a popular material due to its excellent properties. One of the intriguing properties is that carriers in graphene possess relativistic-like massless fermionic feature, instead of Schrödinger-like fermions [2, 3]. This exotic property has leads graphene as a first material that can create a relativistic-like system in the field of condensed matter. Due to its honeycomb-like lattice having two sublattices in a unit cell, the overlapping among the three nearest neighbor electron fields gives rise to the effective field of electrons in graphene like relativistic massless fermions (m=0) with the energy-momentum relation $E(m=0) = v_F \sqrt{p_x^2 + p_y^2 + (mv_F)^2} = v_F p$ where $v_F = 10^6$ m/s is the Fermi velocity of electron in graphene playing the role of the speed of light. However, since this relativistic band structure resulting from the honeycomb like geometry, it is to say that band structure of graphene depends on its geometry. Fortunately, experiment has suggested that graphene can sustain uniaxial strain beyond 20% [4, 5]. Hence, importantly this is possible to control graphene's electronic property by large mechanical strain. Strongly deformed graphene is also possible as we apply a large strain. Recently, theoretical studies have predicted that large strain in zigzag direction can open the energy gap between valence and conduction band [6, 7]. However, as a result, gap opening in strained graphene can not be accessible in case of

applying strain in armchair direction. The carriers with respect to the influence of strain behave like anisotropic massless fermions or asymmetric Weyl-Dirac fermions, leading to strain-induced anisotropic quantum transport in graphene-based systems [8-10].

In this paper, we investigate the density of state and the ballistic conductance in strongly strained graphene using the Landauer formula. We assume that the carriers in strained graphene are governed only by the $\pi-$electrons. Their motions near the Dirac point is studied based on the tight-binding model. The effect of the $\sigma-$electrons can be neglected due to the fact that they are not close to the Dirac point for the considered regime of strain below the critical value, predicted by the first principle calculation [7]. In general, graphene is a normal conductor, except for case of which it is induced by means of proximity effect. The present work will theoretically show that pure normal graphene may tern into directional-dependent superconductor as it is under a heavier strain.

## 2. Model of electrons in deformed graphene

We begin with a usually tight-binding-based Hamiltonian to describe the motions of quasiparticles in strained graphene as given by [6-8]

$$H = \begin{pmatrix} 0 & -\sum_{s=1}^{3} t_s e^{-i\vec{k}.\vec{\sigma}_s} \\ -\sum_{s=1}^{3} t_s e^{i\vec{k}.\vec{\sigma}_s} & 0 \end{pmatrix}. \quad (1)$$

Here, as related to Fig.1, since strain **S** is applied into zigzag direction, the hoping integrals in strained graphene can be defined as $t_1 = t_2 = t$ and $t_3 = \eta t$ with $t_{1,2,3} \sim t_0 e^{-3.37(\frac{|\vec{\sigma}_{1,2,3}|}{c}-1)}$, and the three displacement vectors are therefore obtained as $\vec{\sigma}_{1(2)} = c\langle -(+)(\sqrt{3}/2)(1+S), (1/2)(1-pS)\rangle$ and $\vec{\sigma}_3 = c\langle 0, -(1-pS)\rangle$, where $t_0 \cong 2.7$ meV, $c \cong 0.142$ nm and $p \cong 0.165$ [6-8].

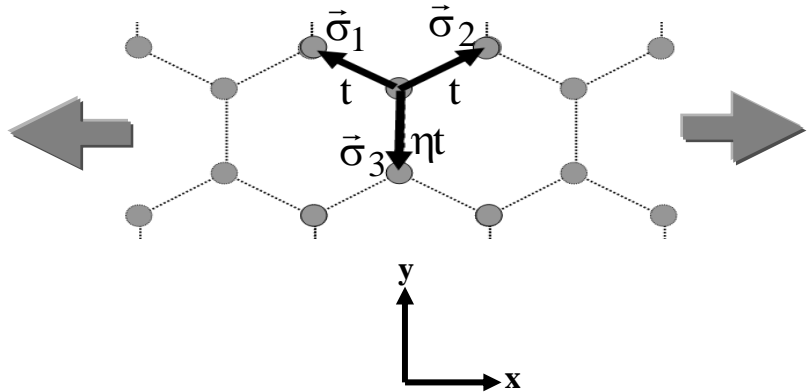

**Figure 1** shows strain dependence of atomic structure of graphene where strain is applied into the zigzag direction or the x-direction.

The Hamiltonian in (1) yields energy dispersion for $\eta < 2$ considered as a case of gapless graphene, as given by [7, 8]

$$E = \hbar\sqrt{(v_x k_x)^2 + (v_y k_y)^2},  \quad (2)$$

where $v_x = 2tL_x\sqrt{1-\dfrac{\eta^2}{4}}/\hbar$, and $v_y = \eta t(L_y + c')/\hbar$

with $L_x = (1+S)c\sqrt{3}/2$, $L_y = (1-pS)c/2$ and $c' = c(1-pS)$.

For this case $S<S_C$, current in graphene are carried by the asymmetric Weyl-Dirac fermions. Note that for $\eta = 2$ or the critical point of the transformation from gapless to gapped graphene, we have approximately strain of $S=S_C \approx 23\%$ [6]. The energy gap will be seen for $\eta > 2$ or $S>S_C$.

**3. Density of state (DOS)**

We calculate the density of state using the usual formula which is a summation over $\vec{k}$ space

$$DOS = 4\sum_{\vec{k}} \delta(E - E(\vec{k})),  \quad (3)$$

where $\delta(E)$ is a Dirac delta function and "4" is stood for two valley and two spin degeneracy. Let us change the summation into integration, as given by

$$DOS = \frac{4WL}{(2\pi)^2}\int_{-\pi}^{\pi}d\theta \int_0^{+\infty} kdk\, \delta(E - E(\vec{k})),  \quad (4)$$

where W and L are width and length of graphene sheet, respectively. From (2) and condition of $k_x = k\cos\theta$ and $k_y = k\sin\theta$, we then have $k = E/\hbar v_\theta$, where $v_\theta = \sqrt{(v_x \cos\theta)^2 + (v_y \sin\theta)^2}$. The region in the integration in (4) needs to change to the new coordinate ie., $\iint_R d\theta dk \to \iint_{R'} d\beta dE$, where $\theta(\beta, E) = \beta$ and $k(\beta, E) = E/\hbar v_{\theta=\beta}$. The equation (4) now takes the form

$$DOS = \frac{WL}{(\pi)^2}\int_{-\pi}^{\pi} d\beta \int_0^{+\infty} k(\beta, E)\begin{vmatrix} \partial\theta/\partial\beta & \partial k/\partial\beta \\ \partial\theta/\partial E & \partial k/\partial E \end{vmatrix} dE\, \delta(E - E(\vec{k})),$$

$$= \frac{WL}{(\pi)^2}\int_{-\pi}^{\pi} \frac{d\beta}{(\hbar v_\beta)^2} \int_0^{+\infty} EdE\, \delta(E - E(\vec{k})).$$

Now, we finally have

$$\frac{\text{DOS}}{\text{DOS(strain = 0)}} = (\int_{-\pi}^{\pi} \frac{d\beta}{2\pi} \left( \frac{1}{(\frac{v_x}{v_F}\cos\beta)^2 + (\frac{v_y}{v_F}\sin\beta)^2} \right)$$

$$= (\int_{-\pi}^{\pi} \frac{d\beta}{2\pi} \left( \frac{1}{(\frac{v_y}{v_F}\cos\beta)^2 + (\frac{v_x}{v_F}\sin\beta)^2} \right).$$

(5)

The density of state for unstrained graphene $\text{DOS(strain = 0)} = \frac{2WLE}{\pi(\hbar v_F)^2}$ is a well known formalism and agrees with our formula (5) for $v_x = v_y = v_F$ when S=0.

Using (2), we will see that for $S \to S_C$, we have $v_x \to 0$ and $v_y \cong v_F$. This leads to the density of state DOS in (5) to tern into

$$\frac{\text{DOS}(S \to S_C)}{\text{DOS(strain = 0)}} \approx \int_{-\pi}^{\pi} \frac{d\beta}{2\pi} \left(\sec^2\beta\right) \approx \frac{2}{\pi}\tan(\pi/2) \to \infty. \quad (6)$$

This result is also plotted in Fig.2. DOS near the critical strain increases sharply to be as an infinite magnitude, representing a new aspect of graphene's density of state.

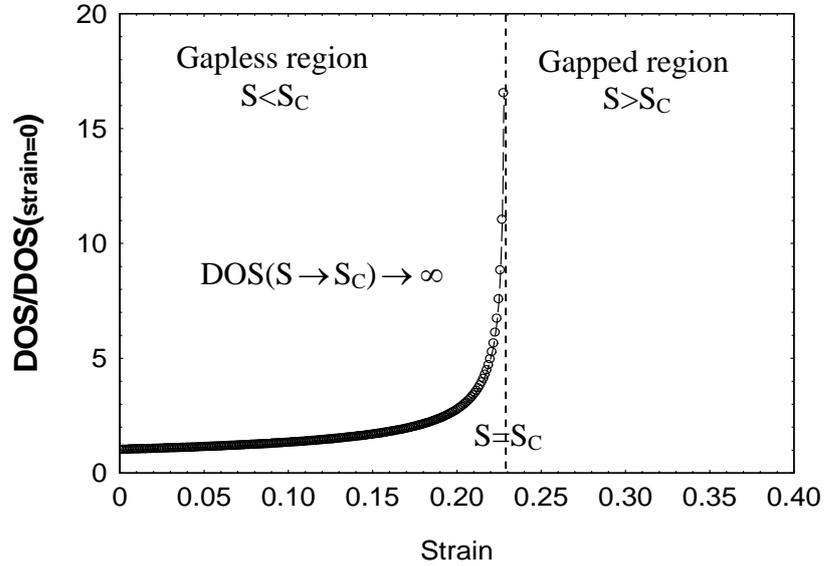

**Figure 2** shows strain dependence of DOS. It sharply increases near the critical strain with respect to the influence of $v_y/v_x \to \infty$ as strain approaches $S_C$ [9].

### 3. Ballistic conductance

In this section, we study the ballistic conductance of strained graphene based on the following formalism, the Landauer formula, as given by

$$G = \frac{2e^2}{h} \int_{-\infty}^{\infty} dE\, T(E) N(E)(-\frac{\partial f}{\partial E}), \quad (7)$$

where the transmission $T(E)=1$ and $-\frac{\partial f}{\partial E} = \frac{e^{(E-E_F)/k_B T}}{(1+e^{(E-E_F)/k_B T})^2}$. Here, $N(E)$ is the number of tunneling mode and it now depends on strain. $N(E)$ is defined by the following formula

$$N(E) = 2\sum_{\vec{k}} \delta(E - E(\vec{k})) \frac{\pi\hbar}{L} \left| \frac{1}{\hbar}(\partial E/\partial k_x) \right|. \quad (8)$$

By following the method similar to that in the section 2, eqn.(8) becomes

$$N(E) = \frac{2WL}{(2\pi)^2} \int_{-\pi}^{\pi} d\theta \int_0^{+\infty} k\, dk\, \delta(E - E(\vec{k})) \frac{\pi\hbar}{L} \left| \frac{1}{\hbar}(\partial E/\partial k_x) \right|. \quad (9)$$

Using the dispersion relation given by (2), the derivative of energy with respect to the wave vector in the x-direction is given by

$$\left| \frac{1}{\hbar} \frac{\partial E}{\partial k_x} \right| = \frac{v_x^2 \cos\theta}{\sqrt{(v_x \cos\theta)^2 + (v_y \sin\theta)^2}}.$$

Let us carry out this by the same way represented in the previous section. The final result of eqn.(9) can be consequently obtained as

$$N(E) = N_x(E) = N_0(E) \left( \frac{1}{2} \int_{-\pi/2}^{\pi/2} d\beta \cos\beta \frac{(\frac{v_x}{v_F})^2}{\left( (\frac{v_x}{v_F}\cos\beta)^2 + (\frac{v_y}{v_F}\sin\beta)^2 \right)^{3/2}} \right),$$

and

$$N_y(E) = N_0(E) \left( \frac{1}{2} \int_{-\pi/2}^{\pi/2} d\beta \cos\beta \frac{(\frac{v_y}{v_F})^2}{\left( (\frac{v_y}{v_F}\cos\beta)^2 + (\frac{v_x}{v_F}\sin\beta)^2 \right)^{3/2}} \right),$$

$$(10)$$

where $N_0(E) = \frac{2WE}{\pi\hbar v_F}$ is a formula associated with the zero-strain limit. The notation x(y) is denoted for the current in x(y) direction or in zigzag (armchair) direction.

To reconsider the conductance formalism in (7), for the zero temperature approximation $-\frac{\partial f}{\partial E} = \delta(E - E_F)$, it will then be reduced into

$$G = \frac{2e^2}{h} N(E_F). \quad (11)$$

As we have seen the result in (10), therefore, the conductance in our model (11) is also directional dependent ie.,

$$G_{x(y)} = \frac{2e^2}{h} N_{x(y)}(E_F). \quad (12)$$

In (12), we will see that $G_x \neq G_y$ for the presence of strain, while $G_x = G_y$ for the absence of strain. Obviously, anisotropic transport property is induced by strain.

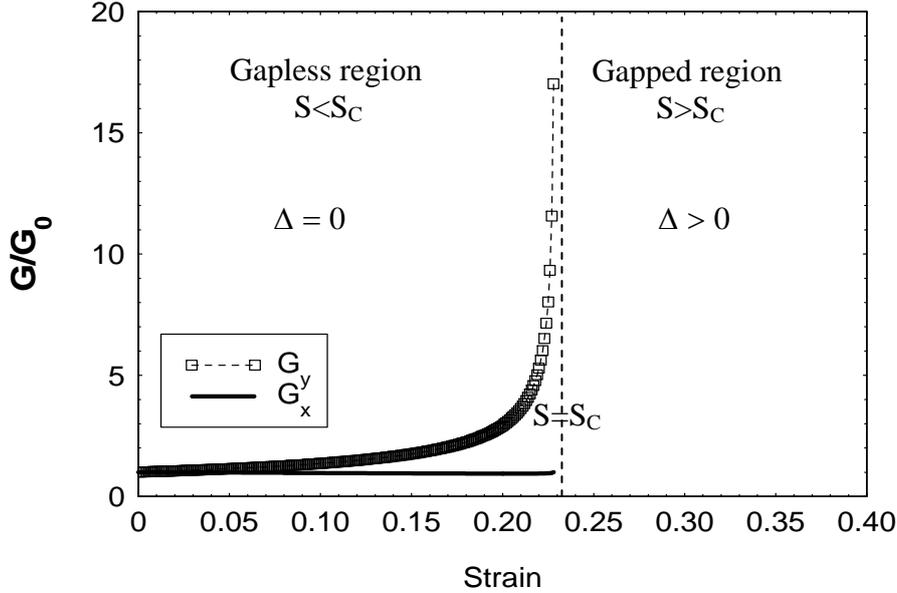

**Figure 3** shows the strain dependence of conductances $G_x$ and $G_y$. At the critical strain $G_y \to \infty$. $\Delta$ denotes the strain-induced energy gap above the critical value.

As a result and discussion of the conductances of the currents in x and the y-directions, our numerical result of the conductances scaled by a unit of conductance in unstrained graphene $G_0 = \frac{2e^2}{h} N_0(E_F)$ is obtained in Fig.3. As an interesting result, $G_x$ is almost independent on strain. In contrast to $G_y$, for strain approaches the critical value $S_C$, it is tremendously large $G \to \infty$. We can analytically obtain $G_y$ based on (10) and (12), as given by

$$G_y(S \to S_c)/G_0 = \int_0^{\pi/2} d\beta \sec^2 \beta \to \infty. \quad (13)$$

The superconductance found near the critical strain in the y-direction is due to the fact that strain-induced anisotropic fermions to carry the current in the system. This behavior should directly result from highly anisotropic fermions $v_y/v_x \gg 1$ found for strain approaching $S_C$. Since we have defined the resistance in the y-direction as of the form $R_y = 1/G_y$ and a unit resistance $R_0 = 1/G_0$, the result for $R_y$ is found to vanish at the critical strain. This is quite similar to the behavior of resistance in superconducting material that may vanish at the critical temperature (see Fig.4).

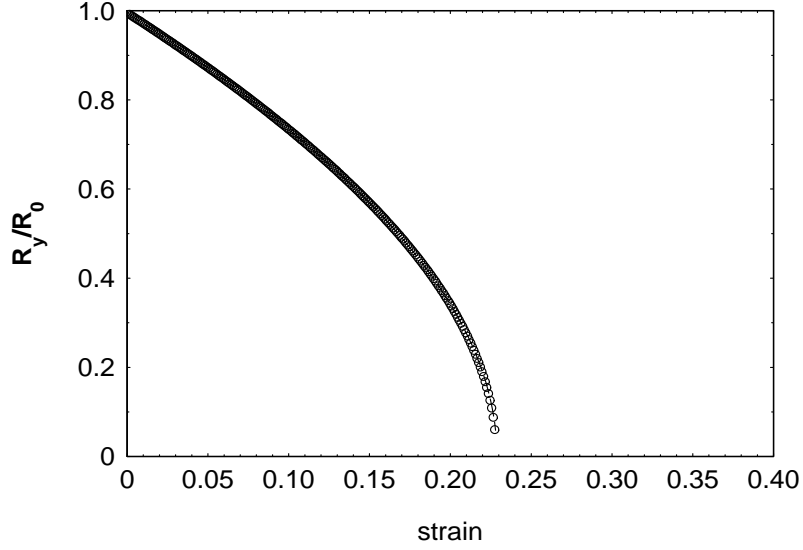

**Figure 4** shows that the resistance in the y-direction vanishes at the critical strain resembling the resistance in superconductor that may vanish at the critical temperature.

## 4. Discussion

As we have partially discussed in the previous section, we found an interesting result that graphene may turn into a directional superconductor. The conductance in ballistic limit is found to be tremendously large at the critical stain, the transition point between gapless and gapped graphene. In this section, we will refer to the Hamiltonian in (1) that may be devised into the three subformulae. The expansion around the Dirac point give rise to equation for each region. For $S<S_C$, Hamiltonian in (1) may be obtained by [7-9, 11]

$$H_i \cong \hbar v_x \sigma_x k_x + \hbar v_y \sigma_y k_y , \qquad (14)$$

where, $v_x$ and $v_y$ are obtained in (2). In the case of applying strain at the critical deformation, Hamiltonian in (1) may be obtained by [12, 13]
for $S=S_C$,

$$H_{ii} \cong \sigma_x \frac{(\hbar k_x)^2}{2m'_x} - \sigma_y (\hbar v'_y k_y) , \qquad (15)$$

where $m'_x = \hbar^2/2tL_x^2$ and $v'_y = (2t/\hbar)(L_y + c')$ [12]. In the gapped region, for $S>S_C$ the Hamiltonian in (1) may be obtained as of the form [14]

$$H_{iii} \cong \sigma_x \left[ \Delta + \frac{(\hbar k_x)^2}{2m'_x} \right] - \sigma_y (\hbar v''_y k_y) , \qquad (16)$$

where $v''_y = \frac{(2\eta - 1)}{3} v'_y$. In the case of strain of $S_C$, we will see the quadratic term appears in the x-direction. However the $H_{ii}$, for $E_F$ being much small, may be approximated as $H_{ii} \to \sigma_x \hbar v'_x k_x - \sigma_y (\hbar v'_y k_y) \cong -\sigma_y (\hbar v'_y k_y)$ similar to $H_i$ at $S \to S_C$. Hence, for the zero limit of the Fermi energy $E_F \to 0$, we may describe electron for $S \leq S_C$ by $H_i$ and the ballistic resistance at $S=S_C$ should vanishes as seen in Fig. 4.

Let us next discuss the fermionic feature for $S>S_C$ based on Hamiltonian (16). For the limit of the zero Fermi energy approximation $E_F \to 0$, we have $k_{x,y} \to 0$. Therefore, $H_{iii}$ becomes of an interesting form which may be obtained by

$$H_{iii} \cong \sigma_x \Delta \to \begin{pmatrix} 0 & \Delta \\ \Delta & 0 \end{pmatrix}. \qquad (17)$$

This is to say that equation of motion of electron for $S>S_C$ resembles the equation of motion of quasiparticles or the Bogoliubov-de Gennes equations in conventional superconducting state where the energy gap of graphene resembles the superconducting gap for order parameter with zero superconducting phase $\Delta(\to 0) = \Delta e^{i\varphi} \to \Delta$, ie.,

$$H_{BdG}(p \to 0, E_F \to 0) = \begin{pmatrix} p^2/2m - E_F & \Delta \\ \Delta & -p^2/2m + E_F \end{pmatrix} \to \begin{pmatrix} 0 & \Delta \\ \Delta & 0 \end{pmatrix}.$$
(18)

Because of this result, we could describe how the resistance in strained graphene evolves into zero resistance like a superconductor at the critical strain.

**5. Summary**

We have investigated density of state and ballistic conductance in graphene under strain where strain is applied in zigzag direction. The strongly asymmetrical velocity resulting from strain causes a superconductance $G \to \infty$ at the critical strain for the current in armchair direction. This predicted result leads to resistance vanishes $R \to 0$ at the critical strain, resembling a transport property of a superconducting material at the critical temperature. This is also obvious that the presence of the strain-induced energy gap above critical strain should possibly be an associated order parameter for such superconducting-like phase transition. Our work has predicted strain-induced directional superconductivity in graphene in which, in this regime, magnitude of strain should be accessible.

**References**
[1] K.S. Novoselov et al., Science **306** (2004) 666.

[2] K. S. Novoselov et al., Nature **438** (2005) 197.

[3] Y. Zhang et al., Nature **438** (2005) 201.

[4] C. Lee et al., Science **321** (2008) 385.
[5] K. S. Kim, et al., Nature London **457** (2009) 706.
[6] V. M. Pereira et al., Phys. Rev. B **80** (2009) 045401.
[7] S.-M. Choi et al., Phys. Rev. B **81**(2010) 081407.
[8] B. Soodchomshom, Physica B **406** (2011) 614.
[9] B. Soodchomshom, J Supercond Nov Magn **24** (2011)1715.
[10] M. Alidoust, J. Linder, Phys. Rev. B **84** (2011) 035407.
[11] S.-L. Zhu et al., Phys. Rev. Lett. **98** (2007) 260402.
[12] B. Soodchomshom, arXiv:1009.5060v1 (unpublished).
[13] P. Dietl et al., Phys. Rev. Lett. **100** (2008) 236405.
[14] O. B.-Treidel et al., Phys. Rev. Lett. **104** (2010) 063901.